\documentstyle[aps,multicol,epsf,epsfig]{revtex}
\begin{document}
\draft
\title{Probabilistic manipulation of entangled photons}
\author{Masato Koashi, Takashi Yamamoto,
 and Nobuyuki Imoto}
\address{CREST Research Team for Interacting Carrier Electronics,
School of
Advanced Sciences, \\
The Graduate University for Advanced Studies (SOKEN),
Hayama, Kanagawa, 240-0193, Japan}
%\date{19 Jul 2000}
\maketitle
\begin{abstract}
We propose probabilistic controlled-NOT and controlled-phase gates
for qubits stored in the polarization of photons. The gates are composed
of linear optics and photon detectors, and consume 
polarization entangled photon pairs. 
The fraction of the successful operation is only limited by 
the efficiency of the Bell-state measurement. 
The gates work correctly under the use of 
imperfect detectors and lossy transmission of photons.
Combined with
single-qubit gates, they can be used for producing arbitrary polarization
states and for  designing various quantum measurements.
\end{abstract}

\begin{multicols}{2}

\narrowtext

Entanglement plays an important role in various schemes of 
quantum information processing, such as  
quantum teleportation \cite{Bennett93}, quantum dense coding
\cite{Bennett92}, certain types of quantum key distributions
\cite{Ekert91}, and quantum secret sharing\cite{Hillery99}. 
It is natural to expect that entanglement shared among 
many particles will be useful for more complicated applications 
including communication among many users. 
Among the physical systems that can be prepared in entangled
states, photons are particularly suited for such 
applications because they can  easily be transferred 
to remote places.  
Several schemes for creating multiparticle entanglement 
from a resource of lower numbers of entangled particles 
have been proposed\cite{anton97,bose98}, 
and experimentally a three-particle entangled state 
[a Greenberger-Horne-Zeilinger (GHZ) state]
was created from two entangled photon pairs \cite{dik99}.

In order to synthesize {\it any} states of $n$ photons
on demand, the concept of quantum gates is useful. 
The universality of the set of the controlled-NOT gate 
and single-qubit gates\cite{sleator95,barenco95} 
implies that you can 
create any states by making a quantum circuit using such gates. 
In addition to synthesizing quantum states, this scheme 
also enables general transformation 
and generalized measurement in the Hilbert space of 
$n$ photons. 
A difficulty in this strategy is how to make two photons 
interact with each other and realize two-qubit gates. 
One way to accomplish this is to implement 
conditional dynamics at the single-photon level
through the strong coupling to the matter such as an atom, and 
a demonstration has been reported \cite{turchette95},
which is a significant step toward 
this goal.
On the other hand, if we restrict our tools to linear optical elements,
a never-failing controlled-NOT gate is impossible, which is 
implied by the no-go theorem for Bell-state measurements
\cite{lutkenhaus99}.  It is, however, still possible to construct a
``probabilistic gate'', which tells us whether the operation
has been successful or not and do the desired 
operation faithfully for the successful cases.
While the probabilistic nature hinders the use for 
the fast calculation
of classical data that outruns classical computers, 
such a gate will still be a useful tool
for the manipulation of quantum states of a modest number of photons,
because no classical computer can be a substitute for this purpose.

In this Rapid Communication, we propose probabilistic two-qubit gates 
for qubits stored in the polarization of photons. The gates 
are composed of photon detectors and 
linear optical components such as beam splitters and wave plates.
As resources, the gates consume entangled photon pairs.
When the detectors with quantum efficiency $\eta$ are used, 
the success probability of $\eta^4/4$ can be obtained, 
which is only limited by the efficiency of the Bell measurement
used in the scheme. 
Combined with single-qubit gates that are easily 
implemented by linear optics, the proposed gates can 
build quantum circuits conducting arbitrary unitary operations
with nonzero success probabilities. 

\begin{figure}
\centerline {\epsfig{width=8.0cm,file=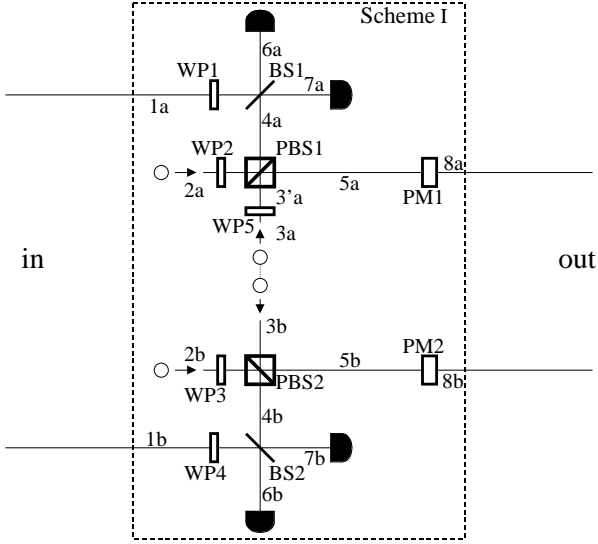}}
\caption{Schematic of the setup of a controlled-phase gate (scheme I).}{
\label{f1}
}
\end{figure}

In Fig.~\ref{f1} we show the schematic of scheme I, 
 the simplest of the schemes we propose in
this paper.
The gate requires two photons and a pair of photons in a Bell
state as resources. Initially, they are in the states
$|H\rangle_{2a}$, $|H\rangle_{2b}$, and
$(|H\rangle_{3a}|V\rangle_{3b}-|V\rangle_{3a}|H\rangle_{3b})/\sqrt{2}$.
The wave plate WP5 rotates the polarization of mode $3a$ by $45^\circ$,
namely, $|H\rangle_{3a}\rightarrow
(|H\rangle_{3^\prime a}+|V\rangle_{3^\prime a})/\sqrt{2}$ and 
$|V\rangle_{3a}\rightarrow
(|H\rangle_{3^\prime a}-|V\rangle_{3^\prime a})/\sqrt{2}$.
After WP5, the entangled photon pair becomes
\begin{eqnarray}
\frac{1}{2}&&(
|V\rangle_{3^\prime a}|V\rangle_{3b}
+|H\rangle_{3^\prime a}|V\rangle_{3b}
\nonumber \\
&&+|V\rangle_{3^\prime a}|H\rangle_{3b}
-|H\rangle_{3^\prime a}|H\rangle_{3b}
).
\end{eqnarray}
The polarizing beam splitter PBS1 transmits $H$ photons and 
reflects $V$ photons. Combined with WP2, it gives the transformation
$|H\rangle_{2a}|H\rangle_{3^\prime a}
\rightarrow 
(|H\rangle_{4a}|H\rangle_{5a}+|V\rangle_{4a}|H\rangle_{4a})\sqrt{2}$
and 
$|H\rangle_{2a}|V\rangle_{3^\prime a}
\rightarrow 
(|V\rangle_{4a}|V\rangle_{5a}+|H\rangle_{5a}|V\rangle_{5a})\sqrt{2}$.
Photons in the $b$ modes are similarly transformed, and we obtain
the state 
\begin{eqnarray}
&&(\beta/2)\left(
|V\rangle_{4a}|V\rangle_{4b}|V\rangle_{5a}|V\rangle_{5b}
+|H\rangle_{4a}|V\rangle_{4b}|H\rangle_{5a}|V\rangle_{5b}
\right.
\nonumber \\
&&\left.
+|V\rangle_{4a}|H\rangle_{4b}|V\rangle_{5a}|H\rangle_{5b}
-|H\rangle_{4a}|H\rangle_{4b}|H\rangle_{5a}|H\rangle_{5b}
\right)
\nonumber \\
&&+\sqrt{1-\beta^2}|\phi\rangle
\equiv \beta|\Psi\rangle+\sqrt{1-\beta^2}|\phi\rangle
,
\end{eqnarray}
where $\beta=1/2$, and $|\phi\rangle$ is a normalized state in which 
the number of photons in mode $4a$ or mode $4b$ is 
not unity.

The photon in the input mode $1a$ $(1b)$ passes through 
wave plate WP1 (WP4), which rotates its polarization by $90^\circ$,
and is mixed with 
the photon in mode $4a$ $(4b)$ by a 50/50 polarization-independent
beam splitter BS1 (BS2). After the beam splitters, the photon number 
of each
mode and each polarization is measured by a photon counter. 
Let us assume
that the state of the input qubits is
\begin{eqnarray}
\alpha_1&&|V\rangle_{1a}|V\rangle_{1b}
+\alpha_2|V\rangle_{1a}|H\rangle_{1b}
\nonumber \\
&&+\alpha_3|H\rangle_{1a}|V\rangle_{1b}
+\alpha_4|H\rangle_{1a}|H\rangle_{1b}.
\label{input}
\end{eqnarray}
The total state after the beam splitter can be calculated 
straightforwardly, but we do not write down the whole
since it is too lengthy. We focus on the terms in which
one $H$ photon is found in modes $6a$ 
or $7a$, one $V$ photon is found in modes $6a$ or $7a$,
and similar conditions hold for $b$ modes. There are 16 
such combinations. For example,
the terms including 
$|V\rangle_{6a}|H\rangle_{7a}|V\rangle_{6b}|H\rangle_{7b}$
are found to be 
\begin{eqnarray}
-&&\frac{\beta}{8}|V\rangle_{6a}|H\rangle_{7a}|V\rangle_{6b}|H\rangle_{7b}
(-\alpha_1|V\rangle_{5a}|V\rangle_{5b}
\nonumber \\
&&+\alpha_2|V\rangle_{5a}|H\rangle_{5b}
+\alpha_3|H\rangle_{5a}|V\rangle_{5b}
+\alpha_4|H\rangle_{5a}|H\rangle_{5b}),
\end{eqnarray}
and the terms including 
$|V\rangle_{6a}|H\rangle_{6a}|V\rangle_{7b}|H\rangle_{7b}$
are
\begin{eqnarray}
\frac{\beta}{8}&&|V\rangle_{6a}|H\rangle_{6a}|V\rangle_{7b}|H\rangle_{7b}
(-\alpha_1|V\rangle_{5a}|V\rangle_{5b}
\nonumber \\
&&-\alpha_2|V\rangle_{5a}|H\rangle_{5b}
-\alpha_3|H\rangle_{5a}|V\rangle_{5b}
+\alpha_4|H\rangle_{5a}|H\rangle_{5b}).
\end{eqnarray}
As seen in these examples, the state in modes $5a$ and $5b$ depends on 
the photon distribution in modes 6 and 7. However, it is easy to 
check that this dependence is canceled if we introduce a phase shift
by phase modulator PM1, 
$|H\rangle_{5a}\rightarrow |H\rangle_{8a}$ and
$|V\rangle_{5a}\rightarrow -|V\rangle_{8a}$, only for the
cases of
$|V\rangle_{6a}|H\rangle_{6a}$ and $|V\rangle_{7a}|H\rangle_{7a}$, 
and similar operation for PM2. Then, for all 16 combinations, 
the state in modes $8a$ and $8b$ becomes
\begin{eqnarray}
-\alpha_1&&|V\rangle_{8a}|V\rangle_{8b}
+\alpha_2|V\rangle_{8a}|H\rangle_{8b}
\nonumber \\
&&+\alpha_3|H\rangle_{8a}|V\rangle_{8b}
+\alpha_4|H\rangle_{8a}|H\rangle_{8b}.
\label{output}
\end{eqnarray}
The evolution from Eq.~(\ref{input}) to Eq.~(\ref{output}) shows 
that this scheme operates as a controlled-phase gate if 
we assign $|0\rangle=|H\rangle$ and $|1\rangle=|V\rangle$.
The probability of obtaining these results is $\beta^2/4=1/16$.
The factor of $1/4$ appearing here can be understood as 
due to the twofold use of Bell-state measurement schemes
with 50\% success probability, used in the dense coding 
experiment \cite{mattle96}.
If we place two additional wave plates in modes $1b$ and $8b$, which 
rotate polarization by $45^\circ$ and $-45^\circ$, respectively,
we obtain a probabilistic controlled-NOT gate.

Next, we consider the effect of imperfect quantum efficiency 
of photon detectors. In order to characterize the behavior of 
the detector, we introduce the parameter $\eta_2$ in addition 
to the quantum efficiency $\eta$,
in such a way that it detects two photons with 
probability $\eta^2\eta_2$ when two photons simultaneously
arrive. For example, conventional avalanche photodiodes (APDs) 
have $\eta_2=0$ since they cannot distinguish
two-photon events from
one-photon events. 
Use of $N$ conventional
APDs after beam splitting the input to $N$ branches
leads to an effective value of $\eta_2=1-1/N$. Recently, 
a detector with high $\eta$ and 
with clearly distinguishable signals for one- and
two-photon events was also demonstrated\cite{kim99}.

There are two distinctive effects 
 caused by the imperfect quantum efficiency.
The first one is that the detectors report some successful
events as false ones by overlooking incoming photons. The success 
probability of $1/16$ in the ideal case thus reduces to 
$p^{\rm (I)}_{\rm true}\equiv \eta^4/16$. The second effect is 
that the detectors report some failing events 
as successful ones. This may occur when two photons enter mode
$4a$ or $4b$, hence the output mode $8a$ or $8b$ has no photon. After
some simple algebra,  the probability $p^{\rm (I)}_{\rm false}$ of this
occurrence is  obtained as $p^{\rm (I)}_{\rm
false}=\eta^4(3-\kappa)(1-\kappa)/4$, with $\kappa\equiv
\eta(1+\eta_2)/2$. Because of this effect,  after discarding the failing
events indicated by the results of  the photon detection,  
the output of the gate still includes errors at probability
$p^{\rm (I)}_{\rm err}\equiv p^{\rm (I)}_{\rm false}/
(p^{\rm (I)}_{\rm true}+p^{\rm (I)}_{\rm false})$. In the following,
we describe two methods for removing these errors.

The first method is the postselection that is applicable 
when every output qubit of the whole
quantum circuit is eventually
measured by photon detectors. As we have seen, the errors in the 
gate always accompany the loss of photons in the output. We also
observe easily that if the input mode $1a$ $(1b)$ is initially in the 
vacuum state, the output mode $8a$ $(8b)$ has no photon whenever  
the detectors show successful outcomes. This implies that 
if one of the gates in the circuit causes errors, at least 
one photon is missing in the final state of the whole circuit. 
The errors
can thus be  discarded by postselecting the events of every detector at
the end of  the circuit registering a photon. This method also works 
when the Bell-state source fails to produce two photons
reliably and emits fewer photons on occasion.

\begin{figure}
\centerline {\epsfig{width=8.0cm,file=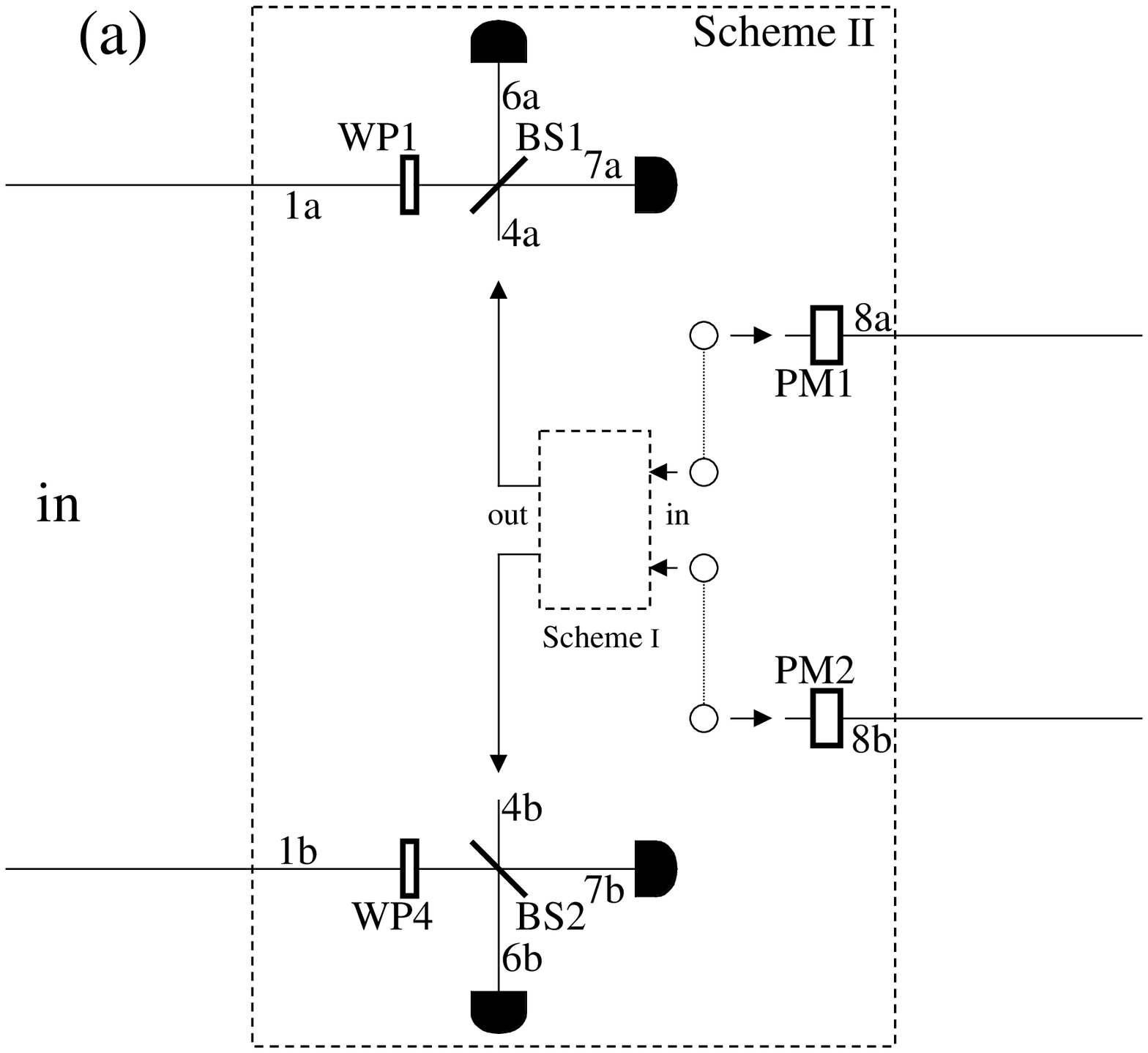}}
\centerline {\epsfig{width=8.0cm,file=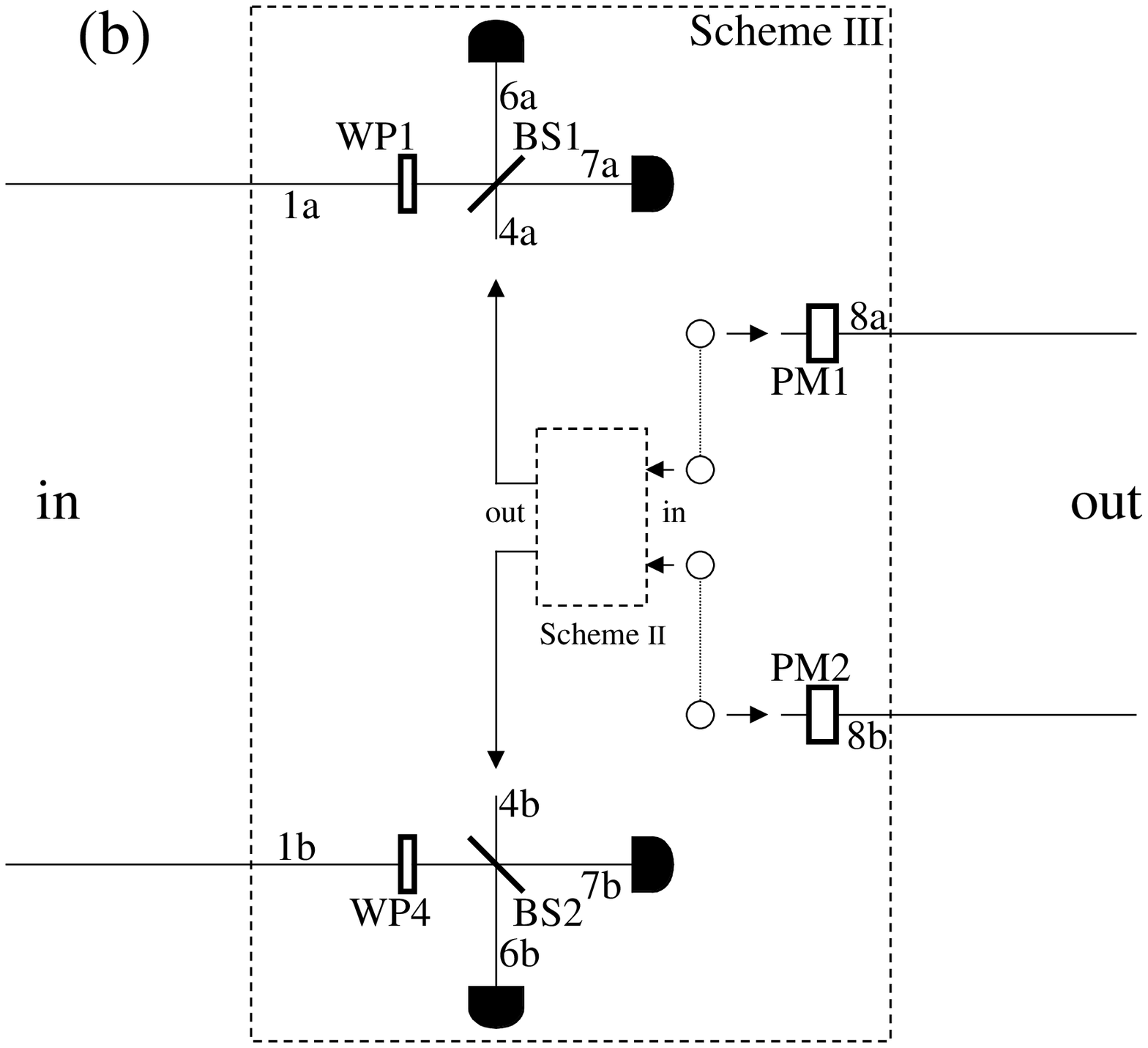}}
\caption{Schematic of the setup of controlled-phase gates.}{
(a) Scheme II. The gate uses scheme I inside. 
(b) Scheme III. The gate uses scheme II inside. 
\label{f2}
}
\end{figure}

The second method is to construct more reliable gates, using 
scheme I for the initialization processes, as shown in Fig.~\ref{f2}(a).
This method is advantageous when the Bell-state source is 
close to ideal and good optical delay lines are available.  
In scheme II, we use two Bell pairs of photons in the state 
$(|H\rangle|H\rangle-|V\rangle|V\rangle)/\sqrt{2}$, and operate
conditional-phase gate (scheme I) on the two photons, one from each
pair. If the operation fails, we discard everything and retry from 
the start. The initialization process is complete when the operation
of scheme I is successful, and the outputs are sent to modes $4a$
and $4b$. The remaining photons of Bell pairs are sent to modes $5a$ 
and $5b$. At this point, the quantum state is prepared in the 
following mixed state,
\begin{equation}
(1-p^{\rm (I)}_{\rm err})
 |\Psi\rangle\langle\Psi |+
p^{\rm (I)}_{\rm err} \hat\rho,
\end{equation}
where $\hat\rho$ is a normalized density operator representing
a state in which no photon exists in mode $4a$ or $4b$.
After this point, the operation of scheme II follows that of 
scheme I. The crucial difference from scheme I is that the
state $\hat\rho$ has no chance to produce successful outcomes.
This leads to $p^{\rm (II)}_{\rm false}=p^{\rm (II)}_{\rm err}=0$,
namely, the faithful operation is obtained with the success probability 
of $p^{\rm (II)}_{\rm true}=(1-p^{\rm (I)}_{\rm err})\eta^4/4$.
When one of the input modes is in the vacuum, this gate never 
reports the successful operation. This implies that  
errors caused by the loss of photons in the upstream circuit
are detected
and hence discarded. 

Using scheme II for the initializing process, we can enhance 
the success probability. Scheme III shown in Fig.~\ref{f2}(b) is exactly
the same as scheme II, except that scheme I inside is replaced by
scheme II itself. When the initialization is completed, the gate inside
produces exactly the state $|\Psi\rangle$. For this scheme, the 
probability of a successful operation is 
$p^{\rm (III)}_{\rm true}=\eta^4/4$, and in the ideal case it is 
$1/4$. This limitation stems from the success probability (50\% each)
of the two Bell-state measurements. 
It should be noted that the maximum of this probability
is still an open question, and if a more efficient way of Bell
measurement is discovered, it will be used in our scheme to enhance
the success probability of the gate.

Scheme III can be viewed as a particular implementation 
of the general scheme of constructing quantum gates using the concept 
of teleportation and Bell measurement\cite{gottesman}, to the case 
of qubits stored in photons. In the general argument that considers 
the use of 
single-qubit gates, three-particle entangled states (GHZ
states) are required as  resources. What was shown here is that  
linear optical components for qubits made of photons 
have more functions than the single-qubit gates, and 
the resource requirement is further reduced to two-particle entanglement.

Since the set of the controlled-NOT gate and single-qubit gates
is universal\cite{sleator95,barenco95}, any unitary transformation
can be realized with a nonzero success probability
 by quantum circuits composed of the proposed gates
and linear optical components. 
For the tasks that take classical data as an input and return
classical data as an output, the quantum circuits here will not 
surpass the conventional classical computers due to 
 the probabilistic nature. But there are other applications 
in which either the input or the output includes quantum states.
For instance, they can be used as a quantum-state synthesizer, which 
produces any quantum state on the polarization degree of freedom with 
a nonzero probability. They are also used as designing various
types of quantum measurement. For any positive operator valued measure
(POVM) \cite{peres} given by the set of positive operators
$\{F_1,\ldots,F_n\}$ with $\sum_k F_k=\bbox{1}$, it is possible to realize
a POVM given  by $\{pF_1, \ldots, pF_n, (1-p)\bbox{1}\}$, where
$p$ is a nonzero probability of success. As transformers of 
quantum states, they may be used for the purification protocol of 
entangled pairs\cite{BBPSSW}. This implies that
if reliable resources of entangled photon pairs are realized,
it may be possible to produce maximally entangled pairs shared 
by remote places connected only by noisy channels.

Finally, we would like to mention the requirement on the 
property of the entangled-pair resources. While the mixing 
of fewer-photon states can be remedied as discussed before,
the mixing of excess photons leads to errors that are difficult
to correct. For example, the photon-pair source by 
parametric down-conversion of coherent light with a pair production
probability of $\eta_{\rm PDC}$ emits two pairs with the 
probability of $O(\eta^2_{\rm PDC})$. This portion 
 causes severe effects when the two or more gates are connected
in series. The recent proposal for the regulated entangled photon pairs
from a quantum dot \cite{benson00} seems to be a promising candidate for
the resources of the proposed gates.

{\it Note added in proof} --- Recently, a proposal of quantum gates for
photons, which is aimed at fast computation, was made by Knill
{\it et al.} \cite{knill}.

This work was supported by a Grant-in-Aid for Encouragement of Young
Scientists (Grant No.~12740243) by the Japan Society of the Promotion of
Science.

\end{multicols}
\end{document}